\documentclass[12pt]{article}
\usepackage{cite,psfig}

\def\be{\begin{equation}}
\def\ee{\end{equation}}
\def\bea{\begin{eqnarray}}
\def\eea{\end{eqnarray}}
\def\xbj{x_{\mbox{\scriptsize Bj}}}
\renewcommand{\epsilon}{\varepsilon}

\newcommand{\mst}[2]{\mbox{\raisebox{-1mm}{$\,\stackrel{#1}{\scriptstyle
#2}\,$}}}

\begin{document}
\thispagestyle{empty}
\vspace*{.5cm}
\noindent
{\hspace*{\fill}\large CERN-TH/2000-306

\vspace{-2mm}
\hspace*{\fill} PITHA 00/26

\vspace{0.7cm}
\hfill October 2000}
\vspace*{2.5cm}

\begin{center}
{\Large\bf Skewed Parton Distributions and $F_2^D$ at $\beta\to 1$}
\\[2.5cm]
{\large A. Hebecker $^a$ and T. Teubner $^b$
}\\[.5cm]
{\it $^a$ Theory Division, CERN, CH-1211 Geneva 23, Switzerland
\\[.1cm] 
$^b$ Institut f\"{u}r Theoretische Physik E, RWTH Aachen, D-52056
Aachen, Germany}
\\[3.1cm]

{\bf Abstract}\end{center}
\noindent
We show that the diffractive structure function is perturbatively 
calculable in the domain where the diffractive mass is small but still 
outside the resonance region. In this domain, which can be characterized 
by $\Lambda^2/Q^2\ll 1\!-\!\beta\ll\sqrt{\Lambda^2/Q^2}$, the structure 
function represents a new observable, which is highly sensitive to the 
small-$x$ skewed gluon distribution. Our leading order calculation and 
the estimate of next-to-leading order corrections are consistent with 
available data and demonstrate the potential of more precise data to put 
further constraints on skewing effects.
\vspace*{2cm}
\newpage

\section{Introduction}
During the last few years, the concept of skewed parton distributions 
(see~\cite{spdf} and refs. therein) has attracted much interest. These 
quantities, which are also known as off-diagonal or non-forward parton 
distribution functions, represent an extension of conventional parton 
distributions and allow for a QCD analysis of hard processes where the 
hadronic target is scattered elastically~\cite{ji}. 

In this letter, we focus on the role of the skewed gluon distribution in 
small-mass inclusive diffraction, i.e. the $\beta\to 1$ limit of 
$F_2^D(\beta,\xi,Q^2)$. The diffractive structure function $F_2^D$ 
is defined by the process $\gamma^*p\to p'X$. Our variables are the photon 
virtuality $Q^2$, the invariant mass $M$ of the diffractive system $X$, and 
the two dimensionless quantities $\beta=Q^2/(Q^2+M^2)$ and 
$\xi=x_{I\!\!P}=\xbj/\beta$. 

We propose the measurement of the diffractive cross section in the kinematic 
domain $Q^2\gg M^2\gg \Lambda^2\,$ (where $\Lambda \sim \Lambda_{\rm
  QCD}$ is a hadronic scale) as a new method for the extraction of the 
skewed gluon distribution at small $x$. At the same time we point out that, 
given the skewed gluon distribution, the diffractive cross section in this 
region is perturbatively calculable. This corresponds to the calculation of 
$F_2^D$ for $\beta$ approaching 1 but still outside the resonance region 
$M\sim\Lambda$. We compare our approach with the limited large-$\beta$ data 
available at present and discuss expected next-to-leading oder corrections 
on the basis of the unintegrated gluon distribution.

\section{$F_2^D$ at very large $Q^2$ and $\beta\to 1$}
The following analysis is based on the well-known formulae~\cite{cc} for 
the $q\bar{q}$ pair production by a virtual photon scattering off a hadronic 
target. At lowest order, the energetic photon fluctuates into a $q\bar{q}$ 
pair with quark, antiquark momentum fractions $z$ and $1-z$ before it
reaches the target. For 
a fixed value of $z$, the typical transverse size of the pair is $\sim 
1/\epsilon$, where $\epsilon^2=z(1-z)Q^2+m^2$ with $m$ the mass of the 
quark~\cite{nz}. If $1/\epsilon$ is sufficiently small, the proton target 
can be characterized by its gluon distribution at the scale $\epsilon$, 
as is known from vector meson electroproduction~\cite{rys,bro}. 

The relevant formulae for longitudinal and transverse incoming photons 
producing a pair of quarks with electric charge $e_q$ read 
\be
\xi F_{L,T}^D(\beta,\xi,Q^2)\,=\,\frac{e_q^2\,\alpha_s^2}{6\,b\,Q^2}\,\,
\beta^3\,(2\beta-1)^2\hspace{-0.3cm}\int\limits_{\textstyle z_{min}}^{
\textstyle\hspace*{.4cm}1-z_{min}}\frac{{\rm d}z}{z(1-z)}\left[\xi G(\xi,
\epsilon^2)\right]^2\,f_{L,T}(z)\,, \label{fltd}
\ee
where 
\bea
f_L(z)&=&\left\{1+\frac{r}{1-1/(2\beta)}\right\}^2\quad,\qquad\qquad\qquad r=
\frac{m^2}{z(1-z)Q^2}\,,\\ \nonumber\\
f_T(z)&=&\frac{f_L(z)}{z(1-z)}\left\{[z^2+(1-z)^2]\left(\frac{1}{\beta}-
(1+r)\right)\left(2-\frac{1}{\beta(1+r)}\right)^{-2}+\frac{r}{4}\right\}\,,
\eea
and the kinematical limits of the $z$ integration are determined by 
\be
z_{min}=\frac{1}{2}-\sqrt{\frac{1}{4}-\frac{m^2}{Q^2(1/\beta-1)}}\,\,.
\label{zm}
\ee
The prefactor $1/b$ is due to the $t$ integration performed under the 
assumption of a $t$ dependence $\sim\exp[\,b\,t\,]$.

Equation (\ref{fltd}) is valid in the leading $\log(1/x)$ approximation, where 
the difference between the longitudinal momenta carried by the two exchanged 
gluons is irrelevant~\cite{bl} and the conventional gluon distribution
appears. To take this difference into account, the gluon distribution
has to be replaced by the skewed gluon distribution $H_g$ according to
\be
G(x,\mu^2)=H_g(x,x,\mu^2)\,\,\to\,\,H(x,x',\mu^2)\,,
\ee
where $x$ and $x'$ are the momentum fractions of the proton carried by the 
first and second exchanged gluons. With the main contribution coming from 
the region $x'\ll x$, the resulting correction can be accounted for by 
introducing the multiplicative factor $R_g=H_g(x,x'\ll x,\mu^2)/ 
H_g(x,x,\mu^2)$ into Eq.~(1) (cf.~the analyses of~\cite{mr,mrt4}). This 
skewing factor has been calculated for $x \ll 1$ in~\cite{sgmr} to be
\be
R_g=\frac{2^{2\lambda+3}}{\sqrt{\pi}}\,\frac{\Gamma(\lambda+\frac{
\scriptstyle 5}{\scriptstyle 2})}{\Gamma(\lambda+4)}\,\,,\qquad\mbox{where}
\qquad\lambda=\frac{{\rm d}\,\ln(xG(x,\mu^2))}{{\rm d}\,\ln(1/x)}\,.
\ee

In deriving Eq.~(\ref{fltd}) the underlying amplitudes were assumed to be
purely imaginary. This can only be justified in the leading $\log(1/x)$ 
approximation. Therefore, in addition to skewing, the effect of the real 
parts of these amplitudes has to be taken into account. Under the assumption 
of appropriate crossing properties of the amplitude $T$ and of a purely 
power-like behaviour of its imaginary part (i.e. a power-like behaviour 
of the gluon distribution), 
\be
\mbox{Im}\,T\sim s^{\lambda}\qquad(\mbox{where}\quad s=Q^2/\xbj\,)\,,
\ee
the relation
\be
C\equiv\frac{\mbox{Re}\,T}{\mbox{Im}\,T}=\tan{\left(\lambda\,\frac{\pi} 
{2}\right)}
\ee
can be derived (see, for example,~\cite{bro,rrml}). 

Equation~(\ref{fltd}) with the correction factor $R_g^2(1+C^2)$ evaluated at 
$x=\xi$ and $\mu^2=\epsilon^2$ forms the basis of our leading order analysis. 

Consider first the simple case when $m=0$ and $\beta\to 1$. In this limit,
$f_L(z)=1$ and $z_{min}=0$ so that Eq.~(\ref{fltd}) provides a parameter-free 
prediction for the longitudinal cross section at asymptotically large 
$Q^2$. This prediction is not affected by the endpoints of the $z$ 
integration since the effective anomalous dimension $\gamma$ of the gluon 
distribution in this region, 
\be
xG(x,\mu^2)\sim \left(\mu^2\right)^\gamma\qquad\qquad (\gamma>0)\,\,,
\ee
suppresses the integrand (cf.~\cite{mrt1}). 

The transverse cross section at $m=0$ and $\beta\to 1$ is suppressed with 
respect to the longitudinal one by an explicit factor $\sim(1-\beta)$ 
contained in $f_T(z)$. However, the $z$ integration in the transverse case 
is divergent even for $\gamma>0$. Since no free quarks can be produced, it 
is natural to use a cutoff $z_{min}\sim \Lambda^2/M^2$, which follows from 
Eq.~(\ref{zm}) if $m$ is replaced by a hadronization scale $\sim\Lambda$. 
(Note that for sufficiently large $\beta$ the scale $\epsilon^2\ge 
z_{min}(1\!-\!z_{min})Q^2$ always stays in the perturbative domain.) The 
$z$ integral in $F_T^D$ then gives a result $\sim 1/z_{min}^{1-2\gamma}$ 
leading to an overall ratio
\be
\frac{F_T^D}{F_L^D}\,\sim\,\frac{M^2}{Q^2}\,\left(\frac{M^2}{\Lambda^2}
\right)^{1-2\gamma}\,.
\ee
This ratio is small as long as 
\be
M^4\ll\left[\left(\Lambda^2\right)^{1-2\gamma}\,Q^2\right]^{\frac{
\scriptstyle 1}{\scriptstyle 1-\gamma}}\,,\label{win}
\ee
which, even when combined with the condition $M^2\gg\Lambda^2$, leaves a 
window for perturbative calculability of $F_2^D$. A particularly simple, 
conservative expression for this window is obtained by setting $\gamma=0$ in 
Eq.~(\ref{win}), which on switching to the kinematic variable $\beta$
leads to
\be
\Lambda^2/Q^2\,\,\ll\,\,1-\beta\,\,\ll\sqrt{\Lambda^2/Q^2}\,.
\ee
In this region, the diffractive structure function is dominated by the 
perturbatively calculable, longitudinal, higher-twist contribution, the 
phenomenological importance of which has been emphasized before~\cite{bekw}. 

Contributions associated with the production of $q\bar{q}g$ final states 
are known to be important for $F_2^D$~\cite{glu}. However, they are not 
relevant in the limit $\beta\to 1$. This can be seen by relating the 
$q\bar{q}g$ final state to the diffractive gluon distribution~\cite{h}, 
which is known to fall like $(1-\beta)^2$ at large $\beta$ in different 
QCD-based models~\cite{fall}. 

Thus, we have established a region at large $Q^2$ and $\beta\sim 1$ where 
the diffractive structure function $F_2^D(\beta,\xi,Q^2)$ is perturbatively 
calculable on the basis of the skewed gluon distribution.

\section{$F_2^D$ at realistic values of $Q^2$ and $\beta$}
\begin{figure}[ht]
\begin{center}
\vspace*{.2cm}
\parbox[b]{11cm}{\psfig{width=11cm,file=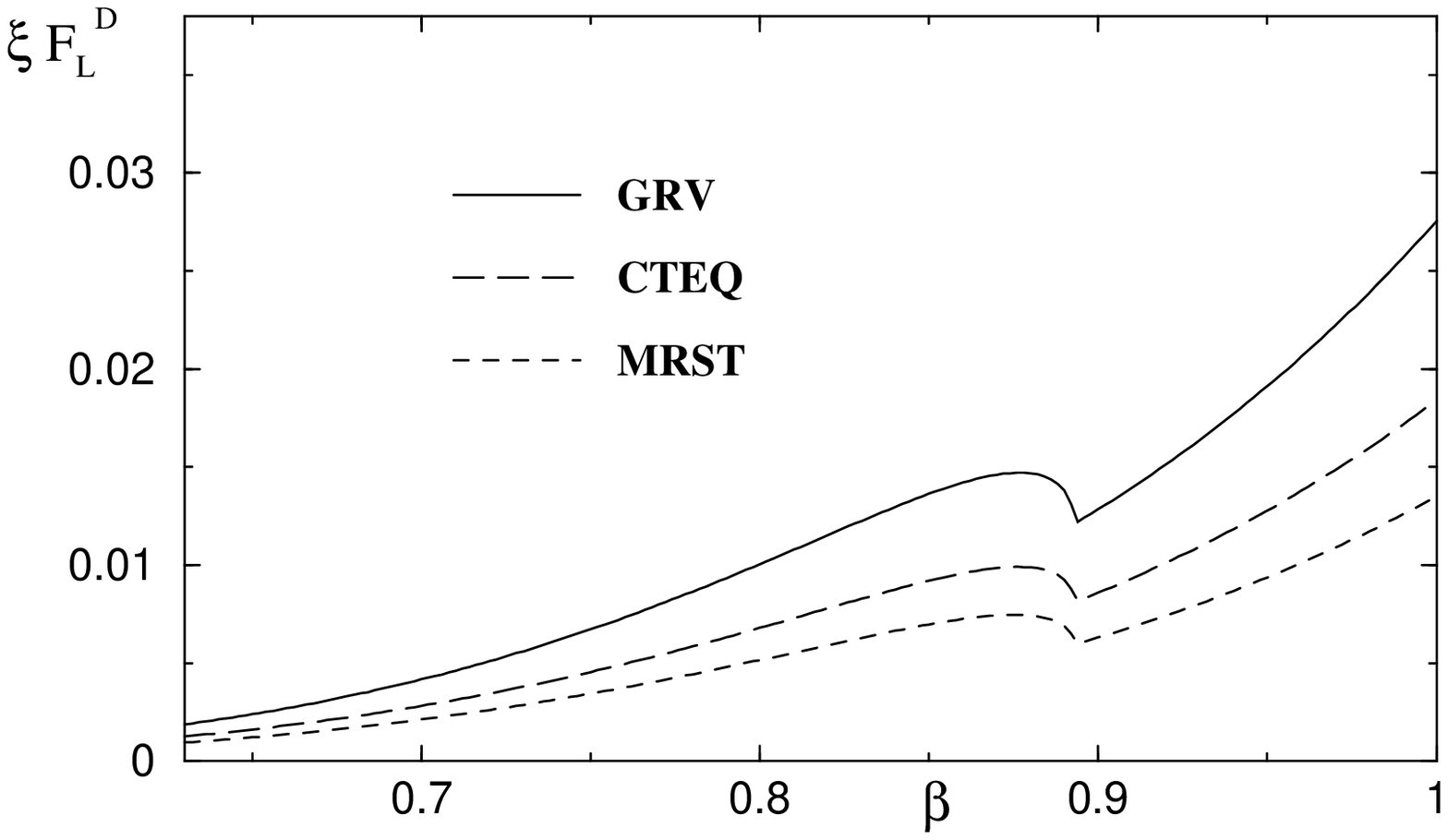}}\\
\end{center}
\vspace*{-0.5cm}
\refstepcounter{figure}\label{comp}
{\bf Figure \ref{comp}:} The longitudinal structure function $F_L^D(\beta,
\xi,Q^2)$ at $Q^2=75~\mbox{GeV}^2$ and $\xi=0.003$ calculated on the basis 
of the leading order GRV~\cite{grv,dur} (grv98lo), CTEQ~\cite{cteq,dur} 
(cteq5l) and MRST~\cite{mrs,dur} (mrs98lo, central gluon) gluon distribution 
functions. The charm threshold is clearly visible. 
\end{figure}

In this section, the diffractive cross section at high $\beta$ and at 
realistic values of $\xi$ and $Q^2$ is calculated on the basis of 
Eq.~(\ref{fltd}), corrected for skewing and the non-zero real parts of the 
amplitudes. Figure~\ref{comp} shows the longitudinal structure function 
obtained with three different parametrizations of the leading order 
gluon distribution. 

In our calculation, we use a charm mass $m_c=1.5$ GeV and a leading order 
running coupling defined by $\Lambda_{\mbox{\scriptsize LO}, n_f=3}=144$ 
MeV ($\alpha_s(M_Z)=0.118$) and evaluated at a fixed scale $\mu^2=Q^2/8$. 
This choice is justified by the variation of $\epsilon^2$, 
which determines the hardness of the process, between 0 and $Q^2/4$ in the 
massless case. Furthermore, we use the slope $b=8$ observed in the closely 
related hard process of elastic $\rho$ meson electroproduction~\cite{slo}. 

In the longitudinal case, no light quark masses are introduced. The 
endpoints of the $z$ integration are cut off by demanding $\epsilon^2>
\epsilon^2_{min}=1.25~\mbox{GeV}^2$ , which is the lowest scale at which 
the MRST gluon distribution is defined. If the cutoff is raised to 
$\epsilon^2_{min}=2.5~\mbox{GeV}^2$, the result changes by $\mst{<}{\sim}5
\%$. This is much less than other theoretical errors, e.g. the considerable 
uncertainty of the small-$x$ gluon distribution (cf.~Fig.~\ref{comp}). 

In the following, we use the CTEQ gluon distribution because of its central 
position in the comparative plot of Fig.~\ref{comp}. For this gluon 
distribution, the effects of skewing and of the real part at $\beta=1$ 
are given by $R_g^2=1.88$ and $1+C^2=1.42$. 

\begin{figure}[t]
\begin{center}
\vspace*{.2cm}
\parbox[b]{11cm}{\psfig{width=11cm,file=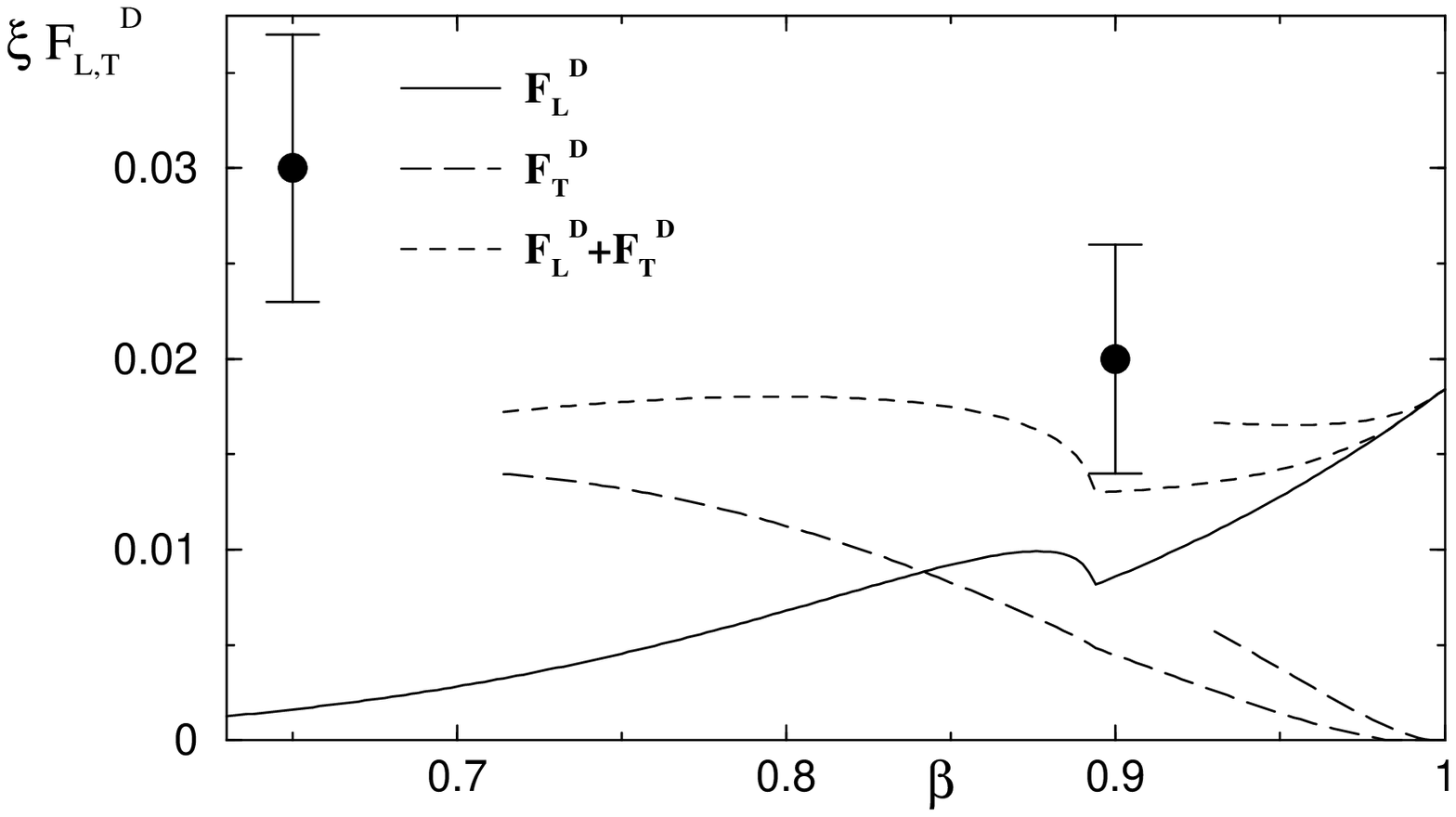}}\\
\end{center}
\vspace*{-0.5cm}
\refstepcounter{figure}\label{h1}
{\bf Figure \ref{h1}:} The diffractive structure functions at $Q^2= 
75~\mbox{GeV}^2$ and $\xi=0.003$ (based on the CTEQ leading order gluon) 
compared with $F_2^D$ data from H1~\cite{data}. The estimates of the 
transverse contribution are based on mass regulators $m_q=0.3$~GeV (upper 
curves) and $m_q=0.6$~GeV (lower curves). 
\end{figure}

Figure~\ref{h1} shows our leading order results together with $F_2^D$ data 
from H1. Unfortunately, at the values of $\beta$ and $Q^2$ available at 
present, the transverse contribution cannot be neglected. Therefore we have 
also included an estimate of the transverse contribution in the plot. This 
contribution was calculated on the basis of Eq.~(\ref{fltd}) with 
skewing and real-part corrections. Unlike the prediction for $F_L^D$,
it is necessary to impose a regulator to estimate $F_T^D$.  Here we
make the $z$ integration finite by giving the light quarks a non-zero
mass $m_q\sim\Lambda$, which is 
justified by the production of massive hadrons in the final state. According 
to Eq.~(\ref{zm}), this mass gives rise to a cutoff $z_{min}$ and therefore 
to a minimal virtuality $\epsilon^2_{min}=m_q^2/(1-\beta)$. At sufficiently 
large $\beta$, this minimal virtuality is in the perturbative domain and 
a calculation on the basis of parametrizations of the gluon distribution 
becomes possible. We have plotted the corresponding curves in the $\beta$ 
range where $\epsilon_{min}^2>1.25~\mbox{GeV}^2$. 

Even though we have argued that $\epsilon^2$ is always in the perturbative 
domain, our transverse cross section should only be interpreted as an 
estimate. The reason for this is the unknown value of the regulator $m_q$ 
or, more generally, the lack of a quantitative understanding of the way in 
which confinement effects in the final state limit the possible kinematics 
of the $q\bar{q}$ pair.

\begin{figure}[ht]
\begin{center}
\vspace*{.2cm}
\parbox[b]{11cm}{\psfig{width=11cm,file=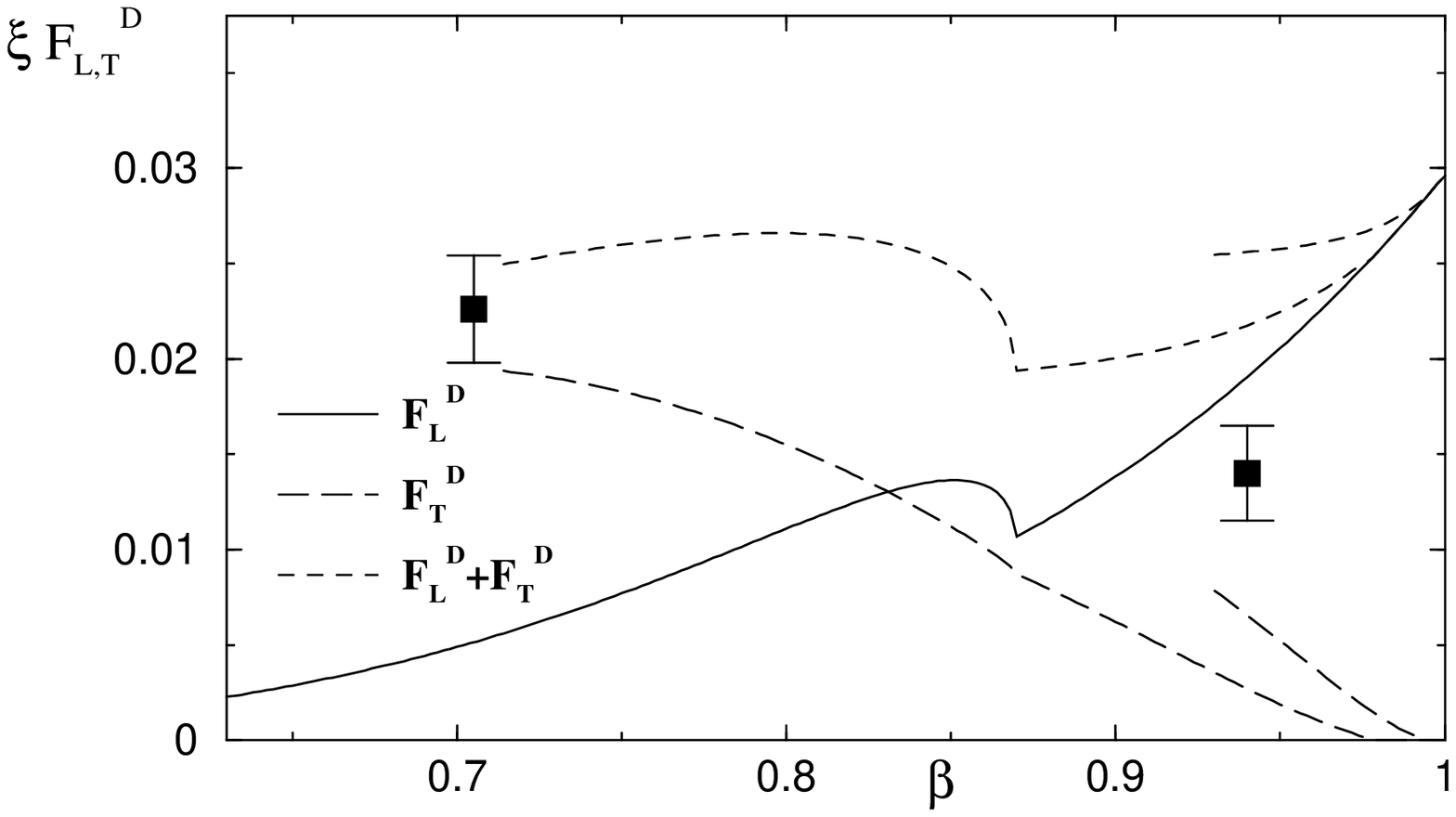}}\\
\end{center}
\vspace*{-0.5cm}
\refstepcounter{figure}\label{zeus}
{\bf Figure \ref{zeus}:} Same as Fig.~\ref{h1}, but for $Q^2= 
60~\mbox{GeV}^2$ and $\xi=0.0042$ with corresponding data from 
ZEUS~\cite{data}.
\end{figure}

In Fig.~\ref{zeus}, the analysis of Fig.~\ref{h1} is repeated for $F_2^D$ 
data from ZEUS. In both figures we have used the data extrapolated by the 
experimental collaborations to fixed $\xi$ and $Q^2$ values, because we are 
interested in the $\beta$ shape at given $\xi$ and $Q^2$. With more data
points and smaller errors one could, in principle, extrapolate the data 
to $\beta\simeq 1$ and thus directly compare with the perturbative $F_L^D$ 
calculation. 

At present, we can only say that the data are approaching the region where 
the calculable longitudinal contribution dominates. Our calculations, which 
include a skewing enhancement of almost a factor 2, are compatible with the 
data. Thus, the prospects for a quantitative description of $F_2^D$ data 
in this domain and for a model-independent test of skewing effects are good.

\section{Beyond leading order}
At present there exists no complete next-to-leading order (NLO) calculation 
of diffractive scattering.  However, it is possible to
improve the predictions beyond the leading approximation.  Above we
have already discussed the important
contributions from the real part of the amplitudes and the effect from
skewed parton distributions.  Both are corrections to the leading
$\log(1/x)$ approximation and lead to a significant enhancement of
$F_2^D$.  Therefore, unless explicitly indicated, they will be taken into
account (on amplitude level, see \cite{mrt4} for details) in all the
numerical results discussed below.  In addition, the simple form of
Eq.~(\ref{fltd}) was obtained
in the leading $\log(Q^2/k_T^2)$ approximation, where $k_T$ is the
transverse momentum of the exchanged gluons that mediate the
diffractive scattering.  Only in this limit can the (loop-)integral over
$k_T$ be performed immediately, leading to the {\em integrated} gluon
distribution.   In the more general case, the amplitudes contain a
non-trivial $k_T$-integration over the {\em unintegrated} gluon
distribution $f(\xi, k_T^2)$,\footnote{For the full amplitudes including
  the integration over the gluon transverse momentum we refer the
  reader to~\cite{lmrt,mrt4}.}  
\begin{equation}
\xi\,G(\xi, Q^2) \ = \ \int^{Q^2} \frac{{\rm d}k_T^2}{k_T^2}\,f(\xi, k_T^2)\,.
\end{equation}
This approach is an application of the so-called $k_T$ or
high-energy factorization~\cite{ktfac} which is an extension of
ordinary collinear factorization.  It was already used in~\cite{lmrt}
for the prediction of diffractive open charm production and found to
increase the cross section considerably.  However, this effect is
nearly compensated by the fact that the NLO gluon parametrizations,
which are used for consistency in our NLO calculations, are smaller
than the leading order ones.

Two additional comments are in order here:  when aiming at a full
NLO prediction, in principle also $q\bar q g$ contributions have to be
taken into account.  While this is certainly true for arbitrary
kinematics, additional real gluon emission is kinematically suppressed for
large $\beta$, which is the case we are interested in.  Virtual gluon
exchange, on the other hand, might be important.  Up to now, such
corrections to the $g g q \bar q$ vertex have not been calculated
yet.\footnote{In a first step the finiteness of the impact factors 
was demonstrated, see~\cite{fm}.}  They might lead to an additional $K$
factor of numerical relevance (see also~\cite{lmrt}).

\begin{figure}[htb]
\begin{center}
\vspace*{-0.5cm}
\parbox[b]{13cm}{\psfig{width=13cm,file=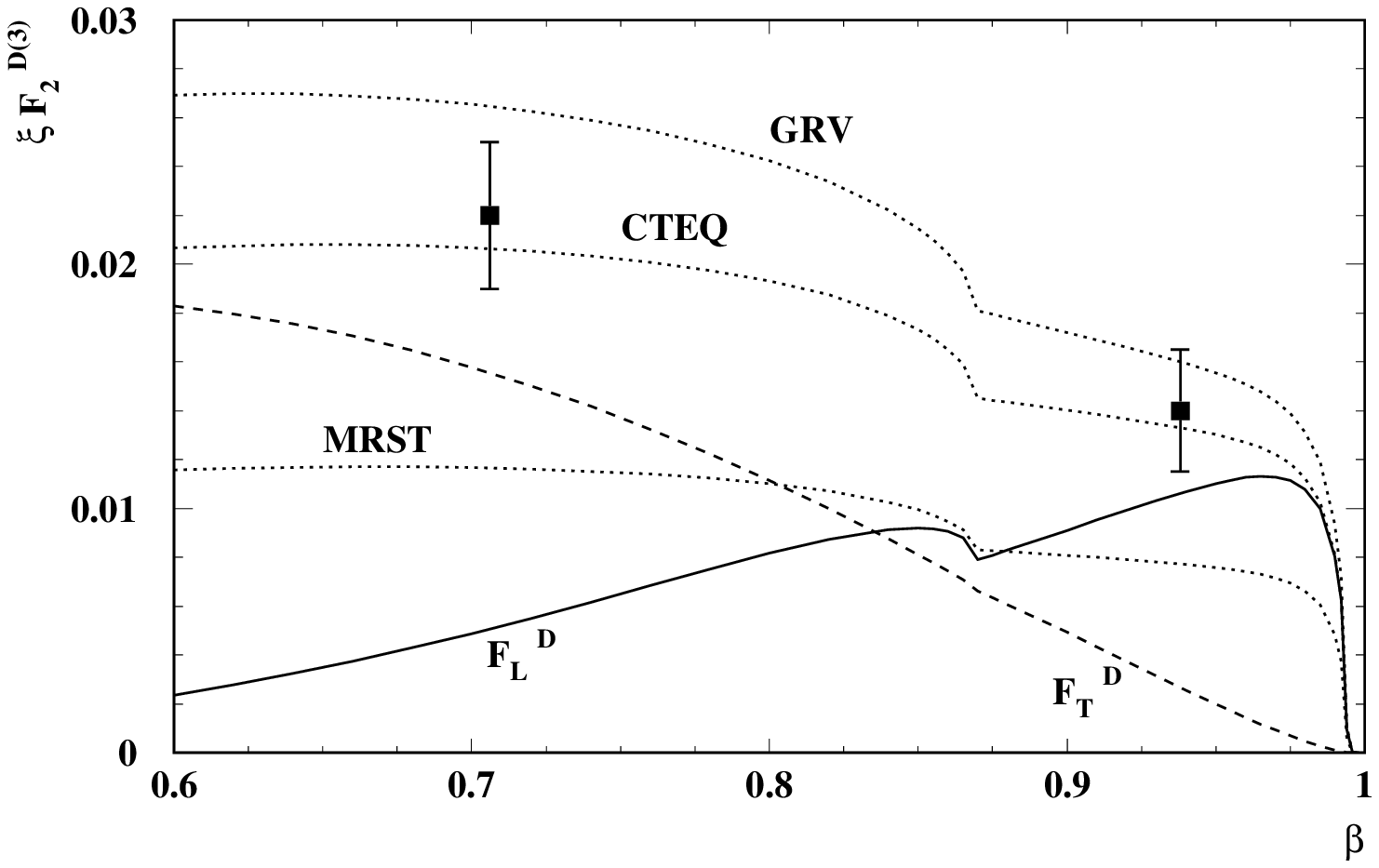}}\\
\end{center}
\vspace*{-1.cm}
\refstepcounter{figure}\label{fig4}
{\bf Figure \ref{fig4}:} $F_2^D$ (at $Q^2 = 60$ GeV$^2$ and $\xi =
0.0042$) in NLO approximation as discussed in the text.  The
continuous (dashed) line shows the longitudinal (transverse)
contribution calculated with the CTEQ5M gluon~\cite{cteq}, the dotted
lines display the sum $F_L^D + F_T^D$ for the three different NLO gluon
parametrizations GRV~\cite{grv} (grv98nlm), CTEQ~\cite{cteq} (cteq5m)
and MRST~\cite{mrs} (mrs99, central gluon) as indicated in the plot.
ZEUS data~\cite{data} are shown for comparison.
\end{figure}
\begin{figure}[htb]
\begin{center}
\vspace*{-0.5cm}
\parbox[b]{13cm}{\psfig{width=13cm,file=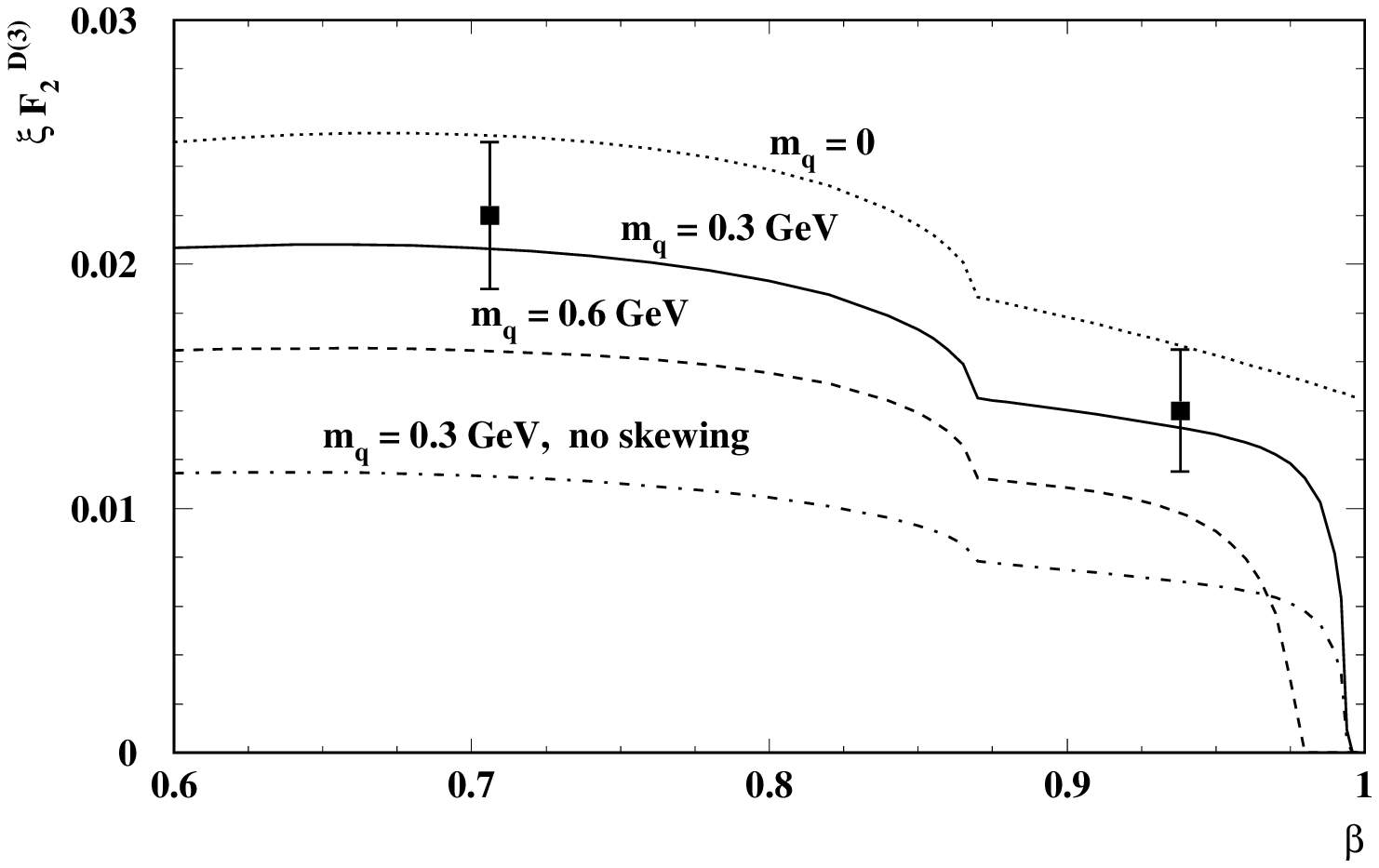}}\\
\end{center}
\vspace*{-1.cm}
\refstepcounter{figure}\label{fig5}
{\bf Figure \ref{fig5}:} $F_2^D$ (at $Q^2 = 60$ GeV$^2$ and $\xi =
0.0042$) for three different choices of the light quark mass as
indicated in the plot.  For comparison two ZEUS data~\cite{data}
points are shown.  The dash-dotted line is the prediction if the
effects from skewing are neglected.
\end{figure}

Let us now discuss the numerical results of our NLO predictions for
$F_2^D$ obtained in the $k_T$-factorization approach. Figure~4 shows
the contributions to $F_2^D$ from $F_L^D$ and $F_T^D$ as continuous
and dashed lines, respectively.  Here $Q^2 = 60$ GeV$^2$, $\xi =
0.0042$, and we have used the CTEQ5M gluon as input in our
calculations.  The three dotted curves correspond to
three different gluon parametrizations, as indicated in the plot; they 
demonstrate the strong dependence of our results on the gluon
distribution used as input.  For comparison we have shown two ZEUS
data points, which are well described by the CTEQ5M prediction.
We should, however, keep it in mind that our calculation is plagued by an
uncertainty from the IR regime.  The size of this uncertainty can be
estimated by varying the light quark mass, which serves as an IR
cutoff.  Figure~5 shows predictions for $F_2^D$ (using the CTEQ5M gluon)
for the three different choices $m_q = 0/0.3/0.6$ GeV.\footnote{Here we can
  calculate $F_2^D$ even for a vanishing quark mass.  This is because no IR
  divergences appear in the integrals over the unintegrated gluon
  distribution, which is continued linearly for very small scales
  $k_T^2 < 1.5$ GeV$^2$ as described in~\cite{lmrt}.} 
(In Fig. 4 $m_q
= 0.3$ GeV was taken as a default value.)  Note that a cutoff given by
the requirement $\epsilon^2 + \kappa_t^2 > (0.3\,\mbox{GeV})^2$, where 
$\kappa_t$ is the transverse momentum of the produced quarks, together 
with $m_q = 0$ leads to results only slightly smaller than the upper curve 
of Fig. 5, obtained with $m_q = 0$ and no other `confinement cutoff'.

As long as (trivial) phase
space suppression is not dominant, the variation of $m_q$ affects mainly
the overall normalization.  Together with the unknown $K$ factor
mentioned above this uncertainty at present makes it difficult to
`extract' the skewed gluon distribution with high accuracy.
Therefore, precise data at higher $Q^2$ and the calculation of the
$K$ factor are highly desirable.  Nevertheless the strong
sensitivity of $F_2^D$ on skewing effects is clearly demonstrated in
Fig.~5 by the dash-dotted curve, which does not include skewing and
has to be compared with the continuous line.  Also note that the enhancement
due to the real part contributions amounts to roughly 40 per cent, in
good agreement with the enhancement factor found in the leading order
analysis.

\section{Conclusions}
In the present letter, we have suggested a new way of extracting the 
small-$x$ skewed gluon distribution from diffractive electroproduction 
data.  We have demonstrated that inclusive diffraction in the kinematic 
domain $Q^2 \gg M^2 \gg \Lambda^2$ is, in principle, a perturbatively 
calculable quantity, which is highly sensitive to skewing effects.  Our 
leading order numerical analysis, which includes a skewing factor $\mst{<}
{\sim}2$, is consistent with $F_2^D$ data now available in the 
region of interest.  Next-to-leading order effects have been estimated 
using the unintegrated gluon distribution.  The results reinforce the 
leading order situation since the enhancement that goes with the explicit 
integration over transverse gluon momenta is largely compensated by the 
smaller next-to-leading order gluon distribution.  It is clear that
future, more precise data have the potential to constrain the small-$x$ 
skewed gluon distribution significantly. 

Our method has the advantage that, unlike the case of diffractive vector
meson electroproduction, no effects of non-perturbative final state
wave functions affect the theoretical prediction.  Furthermore, our 
calculation of inclusive diffraction at $\beta \to 1$ can be used to 
supplement the familiar leading twist partonic analyses of $F_2^D$ 
with a higher twist correction, which is known to be numerically important. 
Thus, we expect that the specific kinematic domain of $F_2^D$ at large 
$Q^2$ and $\beta \to 1$ will play an important role in the future 
development of hard diffraction, both theoretically and experimentally. 

\section*{Acknowledgements}
We would like to thank A.D. Martin and M.G. Ryskin for very helpful 
discussions and comments.


\begin{thebibliography}{99}

\bibitem{spdf}  F.-M. Dittes, D. M\"uller, D. Robaschik, B. Geyer and J. 
                Ho\v{r}ej\v{s}i, Phys. Lett. B209 (1988) 325 and Fortsch. 
                Phys. 42 (1994) 101

\bibitem{ji}    X. Ji, Phys. Rev. Lett. 78 (1997) 610;\\
                A.V. Radyushkin, Phys. Lett. B380 (1996) 417

\bibitem{cc}    M. Genovese, N.N. Nikolaev and B.G. Zakharov, Phys. Lett. 
                B378 (1996) 347;\\
                E.M. Levin, A.D. Martin, M.G. Ryskin and T. Teubner,
                Z. Phys. C74 (1997) 671;\\
                H. Lotter, Phys. Lett. B406 (1997) 171;\\
                M. Diehl, Eur. Phys. J. C1 (1998) 293;\\
                W. Buchm\"uller, M.F. McDermott and A. Hebecker, Phys. Lett. 
                B404 (1997) 353

\bibitem{nz}    N.N. Nikolaev and B.G. Zakharov, Z. Phys. C49 (1991) 607

\bibitem{rys}   M.G. Ryskin, Z. Phys. C57 (1993) 89

\bibitem{bro}   S.J. Brodsky, L. Frankfurt, J.F. Gunion, A.H. Mueller and 
                M. Strikman, Phys. Rev. D50 (1994) 3134

\bibitem{bl}    J. Bartels and M. Loewe, Z. Phys. C12 (1982) 263 

\bibitem{mr}    A.D. Martin and M.G. Ryskin, Phys. Rev. D57 (1998) 6692

\bibitem{mrt4}  A.D. Martin, M.G. Ryskin and T. Teubner, Phys. Rev. D62 
                (2000) 014022

\bibitem{sgmr}  A.G. Shuvaev, K.J. Golec-Biernat, A.D. Martin and 
                M.G. Ryskin, Phys. Rev. D60 (1999) 014015

\bibitem{rrml}  M.G. Ryskin, R.G. Roberts, A.D. Martin and E.M. Levin, Z. 
                Phys. C76 (1997) 231

\bibitem{mrt1}  A.D. Martin, M.G. Ryskin and T. Teubner, Phys. Rev. D56
                (1997) 3007

\bibitem{bekw}  J. Bartels, J. Ellis, H. Kowalski and M. W\"usthoff, 
                Eur. Phys. J. C7 (1999) 443

\bibitem{glu}   N.N. Nikolaev and B.G. Zakharov, Z. Phys. C64 (1994) 631;\\
                H. Abramowicz, L. Frankfurt and M. Strikman, Proc. of the 
                {\it SLAC Summer Institute 1994}, p. 539;\\
                W. Buchm\"uller, M.F. McDermott and A. Hebecker, Nucl. Phys.
                B487 (1997) 283; B500 (1997) 621 (E);\\
                M. W\"usthoff, Phys. Rev. D56 (1997) 4311

\bibitem{h}     A. Hebecker, Nucl. Phys. B505 (1997) 349

\bibitem{fall}  F. Hautmann, Z. Kunszt and D.E. Soper, Phys. Rev. Lett. 81
                (1998) 3333 and Nucl. Phys. B563 (1999) 153;\\
                W. Buchm\"uller, T. Gehrmann and A. Hebecker, Nucl. Phys. 
                B537 (1999) 477

\bibitem{grv}   M. Gl\"uck, E. Reya and A. Vogt, Eur. Phys. J. C5 (1998) 461

\bibitem{cteq}  CTEQ Collab., H.L. Lai et al., Eur. Phys. J. C12 
                (2000) 375

\bibitem{mrs}   A.D. Martin, R.G. Roberts, W.J. Stirling and R.S. Thorne, 
                Eur. Phys. J. C4 (1998) 463; Eur. Phys. J. C14 (2000) 133

\bibitem{dur}   HEPDATA -- The Durham RAL Databases 
                ({\tt http://durpdg.dur.ac.uk/HEPDATA/})

\bibitem{slo}   ZEUS Collab., J. Breitweg et al., Eur. Phys. J. C6 (1999) 
                603

\bibitem{data}  H1 Collab., C. Adloff et al., Z. Phys. C76 (1997) 613;\\
                ZEUS Collab., J. Breitweg et al., Eur. Phys. J. C6 (1999) 43

\bibitem{lmrt}  E.M. Levin, A.D. Martin, M.G. Ryskin and T. Teubner, 
                Z. Phys. C74 (1997) 671

\bibitem{ktfac} S. Catani, M. Ciafaloni and F. Hautmann,
                Phys. Lett. B242 (1990) 97; Nucl. Phys. B366 (1991) 135;\\
                S. Catani and F. Hautmann, Nucl. Phys. B427 (1994) 475;\\
                J.C. Collins and R.K. Ellis, Nucl. Phys. B360 (1991) 3

\bibitem{fm}    V.S. Fadin and A.D. Martin, Phys. Rev. D60 (1999) 114008

\end{thebibliography}
\end{document}